# Multiscale Simulations of Defect Dipole–Enhanced Electromechanical Coupling at Dilute Defect Concentrations

Shi Liu[1] and R. E. Cohen[1,2]

[1]Extreme Materials Initiative, Geophysical Laboratory, Carnegie Institution for Science, Washington DC 20015, USA

[2]Department of Earth and Environmental Sciences, Ludwig-Maximilians-Universität, Munich 80333, Germany

**Abstract**

**The role of defects in solids of mixed ionic-covalent bonds such as ferroelectric oxides is complex. Current understanding of defects on ferroelectric properties at the single-defect level remains mostly at the empirical level, and the detailed atomistic mechanisms for many defect-mediated polarization-switching processes have not been convincingly revealed quantum mechanically. We simulate the polarization–electric field ($P$–$E$) and strain–electric field ($\varepsilon$–$E$) hysteresis loops for $BaTiO_3$ in the presence of generic defect dipoles with large-scale molecular dynamics and provide a detailed atomistic picture of the defect dipole–enhanced electromechanical coupling. We develop a general first-principles-based atomistic model, enabling a quantitative understanding of the relationship between macroscopic ferroelectric properties and dipolar impurities of different orientations, concentrations, and dipole moments. We find that the collective orientation of dipolar defects relative to the external field is the key microscopic structure feature that strongly affects materials hardening/softening and electromechanical coupling. We show that a small concentration (≈0.1 at.%) of defect dipoles dramatically improves electromechanical response. This offers the opportunity to improve the performance of inexpensive polycrystalline ferroelectric ceramics through defect dipole engineering for a range of applications including piezoelectric sensors, actuators, and transducers.**

The presence of defects in ferroelectrics can be the unintended consequence of sample impurities or the result of designed materials engineering.[1] A dipolar defect can be intentionally introduced by substituting a cation with a dopant of different valence. The charge mismatch between the acceptor dopant and the replaced ion is often compensated by nearby oxygen vacancies. Typical acceptor-centered defect complexes include charge-neutral $[(Cu''_{Ti} - V_O^{\cdot\cdot})^\times]$[2, 3] and $[(Mn''_{Ti} - V_O^{\cdot\cdot})^\times]$[4-6] and charged $[(Fe'_{Ti} - V_O^{\cdot\cdot})^{\cdot}]$.[7-9] Donor-doped ferroelectrics are suggested to be charge compensated by the presence of lead vacancies [e.g., $(Nb^{\cdot}_{Ti} - V''_{Pb})'$].[10, 11] Not all dopants form complexes, and their oxidation states depend on the oxygen fugacity of sample growth or annealing.[5] Dipolar defects may couple strongly with the ferroelectric domains/domain walls through long-range electrostatic and elastic interactions, which are suggested to dominate local switching kinetics[12] and appear to be strongly environment/sample dependent. The stabilization of domains/domain walls by the internal bias field arising from the bulk ordering of defect dipoles between an acceptor dopant and an oxygen vacancy is considered to be the origin of domain/domain wall pinning and materials "hardening" (characterized by increased coercive fields),[13-17] and reversible domain switching.[18, 19] The mechanisms of softening by addition of donor dopants are much less understood with several hypotheses proposed, including the compensation of acceptor cations by donor dopants that suppresses the formation of oxygen vacancy,[1] reduced internal stress by lead vacancies that makes domain wall easier to move,[10] and charge transfer between oxygen and lead vacancies that minimizes space charge at domain walls.[20] Recent first-principles investigations of Fe-acceptor-doped and Nb-donor-doped PbTiO$_3$ revealed that the $[(Fe'_{Ti} - V_O^{\cdot\cdot})^{\cdot}]$ defect dipole exhibits a strong tendency to align with the bulk polarization, whereas the $(Nb^{\cdot}_{Ti} - V''_{Pb})'$ shows no binding energy and no preferential alignment with polarization.[21]

Current understanding of defect complexes and their interactions with bulk ferroelectrics remains mostly at the macroscopic/empirical level, and the detailed atomistic mechanisms for many defect-mediated polarization-switching processes have not been established quantum mechanically. Theoretically, first-principles studies have mainly focused on the electronic and static structural properties of point/dipolar defects in doped ferroelectrics[22-24] with a high defect concentration because of the small size of the supercell limited by expensive DFT calculations. Unlike defects in metals and semiconductors where the electrostatic interactions are strongly screened, the dipolar impurities in ferroelectrics couple strongly with the bulk ferroelectrics through the dipolar electric field, resulting in long-range, nonlocal elastic fields.[25] These long-range interactions make it challenging to model and predict the effects of defects on mesoscale ferroelectric properties at finite temperatures. The goal of this work is to develop a quantitative theoretical framework that links the atomic structure of defects and macroscopic ferroelectric properties under realistic doping conditions.

The *P–E* and *ε–E* hysteresis loops contain information on both intrinsic bulk and extrinsic defect-related ferroelectric properties,[26] and decoding intrinsic and extrinsic effects is critical for a complete understanding of defect-mediated polarization switching. We simulated the hysteresis loops in tetragonal $BaTiO_3$ in the presence of dipolar defects using large-scale molecular dynamics simulations (see Methods) under different defect conditions and electric field orientations. A defect dipole is introduced by placing a pair of particles of opposite-charge (*q*) equidistant from a Ti atom. Varying the number, position, and charge of dummy atoms allows the modulation of concentration (*n*), orientation, and magnitude (μ) of dipole impurities, providing a general scheme to study the interplay between dipolar defects and local ferroelectric switching. The samples with randomly oriented defect pairs of zero dipole moment (*n* = 1.2%, *q* = 0.0 *e*, Figure 1a) have essentially the same hysteresis loops as the pure sample (*n* = 0.0 %, Figure S2) in response to electric fields applied along three Cartesian axes ($E_x$, $E_y$, and $E_z$). The *P–E* loops in defect-free and zero-dipole cases possess similar switching fields and symmetric shapes, a signature of isotropic response to electric fields. With the presence of dipolar defects (*n* = 1.2%, *q* = 0.1 *e*, Figure 1b–c), the hysteresis loops acquire drastically different shapes. This highlights the importance of electrostatic interaction between dipolar defects and bulk polarization for modulating ferroelectric switching. Remarkably, the configuration-averaged hysteresis loop for samples with randomly oriented defect dipoles still exhibits a symmetric profile albeit reduced coercive field (Figure 1b). Our results suggest that randomly oriented defect dipoles can reduce the barrier height of ferroelectric double-well potential (materials softening) while preserving the symmetry. This is consistent with softening of ferroelectrics by the addition of donor dopants where defect dipoles such as dopant-Pb-vacancy have little preference to align with the bulk polarization[21] and are therefore more likely to orient randomly. Previous simulations based on phenomenological Ginzburg-Landau theory also showed randomly distributed/oriented defect dipoles could reduce the coercive field.[27]

Aligning dipoles along the −*y* direction causes further changes in hysteresis loops (Figure 1c). Double-hysteresis loops for electric fields ($E_x$ and $E_z$) perpendicular to the collective orientation of defect dipoles appear. The $P_y$–$E_y$ loops shift toward the $+y$ direction, reflecting the presence of a built-in electric field along $-y$. These three $P$–$E$ loops are the natural consequences of "shape memory" effect caused by the aligned dipolar defects. Regardless of the external electric field orientation, once the field is turned off, the sample will converge preferentially to a state with $P_x = P_z = 0$ and $P_y = -0.28$ C/m$^2$, driven by the built-in electric field along $-y$. This is the atomistic manifestation of the "symmetry-conforming principle" proposed by Ren, who reported that diffusionless unswitchable defect dipoles will conform the nearby ferroelectric domains to adopt the same orientation.[18] The spontaneous recovery of original polar/strain state causes giant recoverable strain change (Figure 1c) along $x$ and $z$ directions. It is noted that our force field is parameterized from PBEsol density functional which overestimates the tetragonality of BaTiO$_3$ ($c/a$ = 1.029, the theoretical maximum linear strain change for 90° switching is 2.9% compared to experimental value of ≈0.8% for single-crystal BaTiO$_3$ at 1 MV/cm)[28]. Experimentally, aging ferroelectrics to align defect dipoles could achieve improved electromechanical responses.[19, 29] Our simulations suggest that poling the sample during the aging process is critical in controllably aligning the defect dipoles and applying an electric field perpendicular to the poling direction to realize optimal electro-strain coupling during device operation.

The magnitude of the built-in electric field arising from dipole impurities depends on the concentration and magnitude of defect dipoles. For a collection of dipoles ($q$ = 0.1 $e$) aligned along $-y$, the shape of the $P_x$–$E_x$ loop evolves from square ($n$ = 0.6%) to slim, pinched ($n$ = 0.8%), and eventually becomes double loops ($n$ = 1.2 %) as the concentration increases (Figure 2a). Similar hysteresis loop transition is obtained by increasing the dipole moment ($q$ = 0.025, 0.05, 0.1$e$) at a fixed defect concentration ($n$ = 1.2 %, Figure 2b). The gradual shape change in $P_x$–$E_x$ loops effectively corresponds to the horizontal shift of $P_y$–$E_y$ loops showing increasing internal electric field along $-y$. The "square-slim-double loop" transition commonly occurs with aging of ferroelectric ceramics.[26, 30] Our simulations clearly demonstrate that the internal electric field associated with ordered defect dipoles is responsible for the increasing constriction of ferroelectric hysteresis loops on aging.

We perform first-principles density functional theory (DFT) calculations to relate the generic defect dipole model used in classical molecular dynamics to realistic doping conditions. We first evaluate the effect of dipole impurity on the macroscopic polarization of tetragonal BaTiO$_3$. DFT calculations confirm that the presence of dipolar defects ($n$ = 0.7 % atomic) increases the total polarization along the defect dipole direction, whereas the polarization enhancement ($\Delta P/P_0$, where $P_0$ is the polarization of pure bulk) turns out to be insensitive to dopant types studied in this work (Mg, Mn, and Zn, see Methods). We then map out the dependence of $\Delta P/P_0$ on the classical dipole moment ($\mu_{MD}$) with molecular dynamics. The quantum mechanical value of $\Delta P/P_0$ is used to determine the value of $\mu_{MD}$ that results in the same polarization enhancement (See Methods in Supporting Information).

With the value of $\mu_{MD}$ derived quantum mechanically, we explore the ferroelectric switching under a realistic doping condition and find that even at a relatively low concentration of 0.1 at.% (64 dipolar defects distributed among 13,824 unit cells of 69,120 atoms, Figure 3a), aligned defect dipoles are still capable of driving the spontaneous strain recovery. The application of an electric field along the $-x$ direction at 40 ps drives the exchange of short and long axes, coupled with a giant change in strain (Figure 3b). As the electric field starts to decrease at 90 ps, the structure gradually recovers to its original strain state with the long axis realigned with the defect dipole direction ($-y$) through a polarization rotation process (Figure 3c). We find from the layer-resolved polarization profiles (in the $xy$ plane) that though the $-x$ electric field aligned the majority of ferroelectric dipoles, the defect dipoles strongly pin the nearby ferroelectric dipoles along $-y$ (Figure 3d), resulting in a polarization with substantial $-y$ component for layers containing dipolar defects (Figure 3e). With a reducing electric field, the layers without defect dipoles (initially with $P_x < 0$, $P_y \sim 0$, $t$ = 90 ps) will evolve to align with defected layers with $P_y < 0$, leading to the "shape memory" effect.[31]

The defect dipole-enhanced electromechanical coupling, which is supported by our simulation results, requires that defect dipoles orient and align with the local fields during the ageing process, but are frozen during electric field cycling at room temperature. Experimental studies suggest that the reorientation of $(Mn_{Ti}'' - V_O^{\cdot\cdot})^\times$ defect dipoles in BaTiO$_3$ requires long time, high thermal energy, and high electric fields and is therefore not likely at room temperature.[6, 18] The very low ageing temperature (~80°C) required to align defect dipoles show that the activation energy of reorientation $\Delta E$ should be very high (this is counterintuitive but is clear if one considers how much a Boltzmann factor exp($-\Delta E/k_B T$) changes from room temperature to 80°C: only a large $\Delta E$ will make the Boltzmann factor substantially different between room temperature and 80°C). The slow reorientation of defect dipoles must be reconciled with the relatively fast oxygen diffusion process observed experimentally.[32, 33] It is likely that the binding energy of the oxygen vacancy to the charged cation defect must be strong enough to hold those oxygen vacancies fixed, while unassociated oxygen vacancies are free to move quickly.

This fully ab initio, multiscale approach that combines quantum mechanical calculations and large-scale molecular dynamics offers an atomistic and quantitative understanding of the effects of dipolar impurities on the electromechanical properties of ferroelectrics, highlighting the potential to optimize inexpensive ferroelectric ceramic materials with high electromechanical coupling efficiency through designed defect engineering. Given that all model parameters can be obtained from first-principles calculations, the multiscale approach can be applied to other doped ferroelectrics, potentially enabling the use of high-throughput methods to screen dopants for improved electromechanical coupling via "defect engineering."

**Supplementary Material** See supplementary material for details of computational methods

**Acknowledgments** We thank the Carnegie Institution for Science, the US Office of Naval Research, and ERC Advanced Grant ToMCaT for support. Computational support was provided by the US DOD from the HPCMO and by the Carnegie Institution for Science through computer time at Memex.

**Author Contributions** SL and REC designed and analyzed the simulation approaches. SL performed the molecular dynamics simulations. Both authors discussed the results and implications of the work and commented on the manuscript at all stages.

**Author Information** The authors declare no competing financial interests. Readers are welcome to comment on the online version of the paper. Correspondence and requests for materials should be addressed to SL (sliu@carnegiescience.edu) and REC (rcohen@carnegiescience.edu).

**Figures**

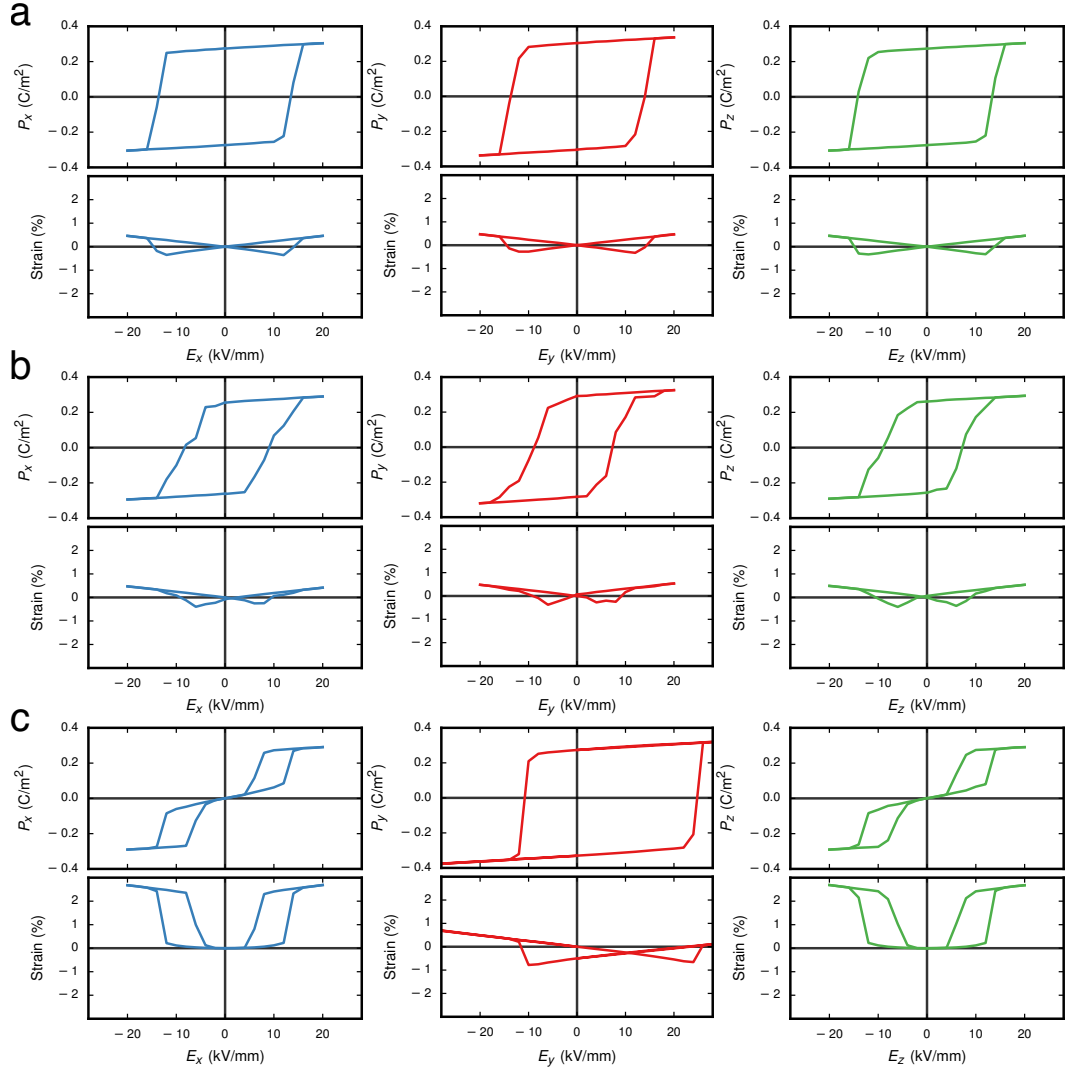

**Figure 1 | Molecular dynamics simulations of polarization–electric field (*P–E*) and strain–electric field (*ε–E*) hysteresis loops in BaTiO₃ with generic defect dipoles.** Electric fields are applied along the *x* (blue), *y* (red), and *z* (green) axes, respectively. For a given defect concentration, the positions of dipolar defects are randomly distributed. The simulated hysteresis loop is averaged over multiple cycles and multiple configurations (see Methods). **a.** Isotropic and symmetric hysteresis loops for $n = 1.2\%$, $q = 0.0\,e$. The loops are essentially the same as those in pure BaTiO₃. **b.** Symmetric hysteresis loops for $n = 1.2\%$, $q = 0.1e$ with defect dipoles randomly oriented. **C.** Double loops along *x* and *z* and horizontally shifted loops along *y* for $n = 1.2\,\%$, $q = 0.1\,e$ with defect dipoles aligned along $-y$.

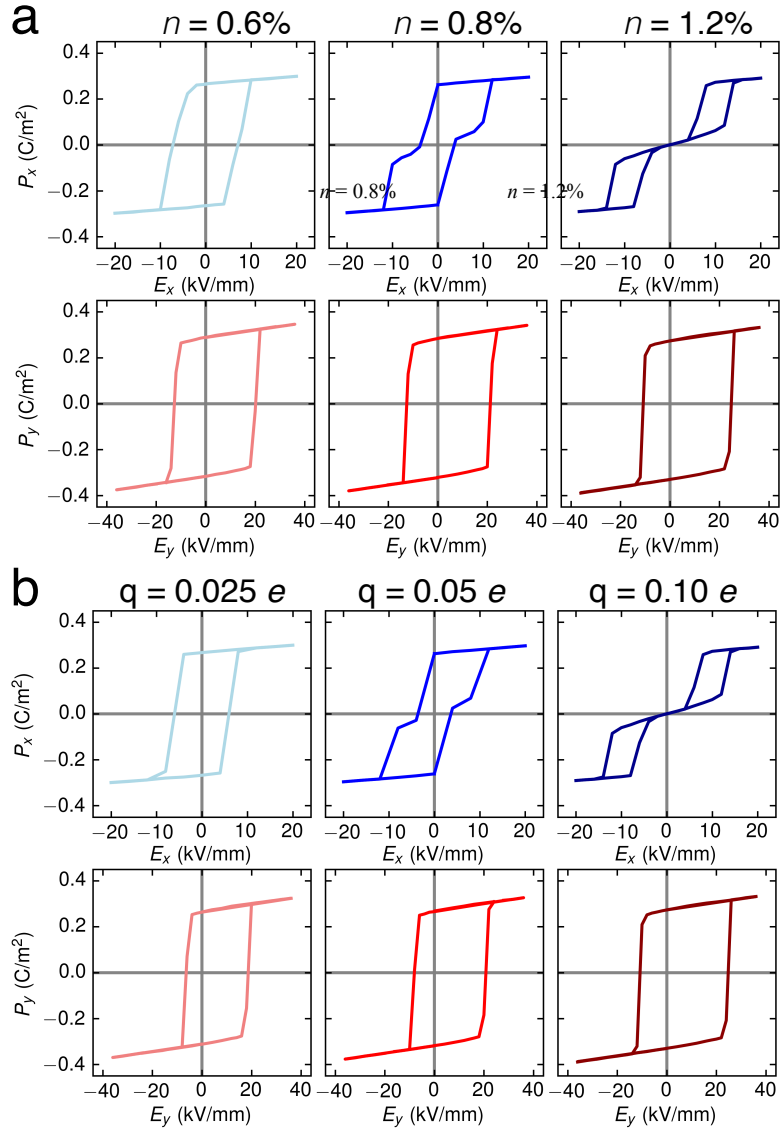

**Figure 2 | Effects of defect concentrations and defect dipole magnitude on hysteresis loops.** **a.** Simulated $E_x$–$P_x$ (top) and $E_y$–$P_y$ (bottom) hysteresis loops for samples with three defect concentrations ($n$ = 0.6%, 0.8%, and 1.0%) for $q$ = 0.1 $e$. **b.** Simulated $E_x$–$P_x$ (top) and $E_y$–$P_y$ (bottom) hysteresis loops for samples with three defect dipole magnitudes ($q$ = 0.025, 0.05, 0.10 $e$) at $n$ = 1.0%. Defect dipoles are aligned along $-y$.

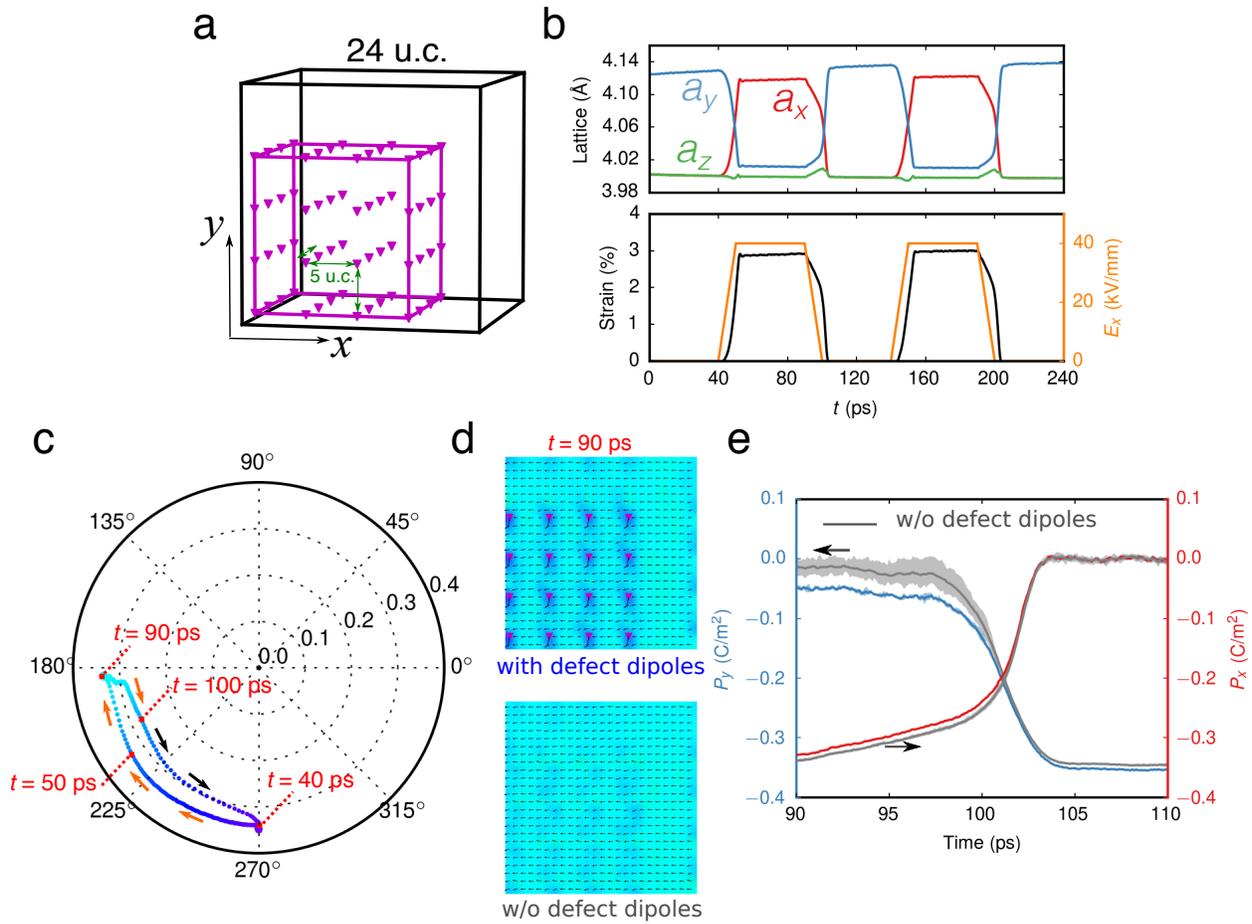

**Figure 3 | Defect dipole–induced reversible electromechanical coupling. a.** A 24 × 24 × 24 supercell with 64 evenly distributed defect dipoles corresponding to a low defect concentration of 0.1 at.%. Neighboring dipolar defects are separated by five unit cells. Simulations using different dipole spacing distances give similar results. Dipolar defects are aligned to maximize the restoring force. **b.** Evolution of lattice constants and strain under pulsed electric fields. The long axis is along −y at $t = 0$ ps. The electric field along −x is turned on at $t = 40$ ps, gradually changing the long axis to −x. At $t = 90$ ps, the electric field starts reducing and is fully turned off at $t = 100$ ps, coupled with a spontaneous strain recovery process with the long axis changing back to the y axis. Applying a second electric field pulse along −x induces another giant strain change. **c.** Polarization rotation in the xy plane under a single pulsed electric field illustrated with polar coordinates. **d.** Polarization profiles for layers with defect dipoles (top) and without defects (bottom) in the presence of the −x electric field ($t = 90$ ps). Black arrows represent local ferroelectric dipoles. **e.** Evolution of layer-resolved polarization for defect layers (blue for $P_y$ and

red for $P_x$) and nondefect layers (gray). Defected layers initially have substantial −*y* polarization, driving the strain recovery process.

**Reference**

1. B. Jaffe, *Piezoelectric ceramics*. (Elsevier, 2012).
2. R.-A. Eichel, H. Kungl and M. J. Hoffmann, J. Appl. Phys. **95** (12), 8092-8096 (2004).
3. R.-A. Eichel, P. Erhart, P. Träskelin, K. Albe, H. Kungl and M. J. Hoffmann, Phys. Rev. Lett. **100** (9), 095504 (2008).
4. R. A. Serway, W. Berlinger, K. A. Müller and R. W. Collins, Phys. Rev. B **16** (11), 4761-4768 (1977).
5. R. A. Maier, T. A. Pomorski, P. M. Lenahan and C. A. Randall, J. Appl. Phys. **118** (16), 164102 (2015).
6. L. Zhang, E. Erdem, X. Ren and R.-A. Eichel, Appl. Phys. Lett. **93** (20), 202901 (2008).
7. H. Meštrić, R.-A. Eichel, K.-P. Dinse, A. Ozarowski, J. v. Tol and L. C. Brunel, J. Appl. Phys. **96** (12), 7440-7444 (2004).
8. H. Meštrić, R. A. Eichel, T. Kloss, K. P. Dinse, S. Laubach, S. Laubach, P. C. Schmidt, K. A. Schönau, M. Knapp and H. Ehrenberg, Phys. Rev. B **71** (13), 134109 (2005).
9. E. Erdem, M. D. Drahus, R.-A. Eichel, H. Kungl, M. J. Hoffmann, A. Ozarowski, J. Van Tol and L. C. Brunel, Functional Materials Letters **1** (01), 7-11 (2008).
10. R. Gerson, J. Appl. Phys. **31** (1), 188-194 (1960).
11. H. Takeuchi, S. Jyomura, E. Yamamoto and Y. Ito, The Journal of the Acoustical Society of America **72** (4), 1114-1120 (1982).
12. P. Gao, C. T. Nelson, J. R. Jokisaari, S. H. Baek, C. W. Bark, Y. Zhang, E. G. Wang, D. G. Schlom, C. B. Eom and X. Q. Pan, Nature Communications **2** (2011).
13. K. Carl and K. Hardtl, Ferroelectrics **17** (1), 473-486 (1977).
14. P. Lambeck and G. Jonker, Ferroelectrics **22** (1), 729-731 (1978).
15. P. Lambeck and G. Jonker, J. Phys. Chem. Solids **47** (5), 453-461 (1986).
16. T. Yang, V. Gopalan, P. Swart and U. Mohideen, Phys. Rev. Lett. **82** (20), 4106 (1999).
17. M. I. Morozov and D. Damjanovic, J. Appl. Phys. **107** (3), 034106 (2010).
18. X. B. Ren, Nat. Mater. **3** (2), 91-94 (2004).
19. Z. Feng and X. Ren, Appl. Phys. Lett. **91** (3), 032904 (2007).
20. L. Eyraud, B. Guiffard, L. Lebrun and D. Guyomar, Ferroelectrics **330** (1), 51-60 (2006).
21. A. Chandrasekaran, D. Damjanovic, N. Setter and N. Marzari, Phys. Rev. B **88** (21), 214116 (2013).
22. P. Erhart, P. Traskelin and K. Albe, Phys. Rev. B **88** (2) (2013).
23. S. Korbel and C. Elsasser, Phys. Rev. B **88** (21) (2013).
24. J. F. Nossa, Naumov, II and R. E. Cohen, Phys. Rev. B **91** (21) (2015).
25. S. V. Kalinin, B. J. Rodriguez, A. Y. Borisevich, A. P. Baddorf, N. Balke, H. J. Chang, L. Q. Chen, S. Choudhury, S. Jesse, P. Maksymovych, M. P. Nikiforov and S. J. Pennycook, Adv. Mater. **22** (3), 314-322 (2010).
26. L. Jin, F. Li and S. J. Zhang, J. Am. Ceram. Soc. **97** (1), 1-27 (2014).
27. R. Ahluwalia and W. Cao, Phys. Rev. B **63** (1), 012103 (2000).
28. E. Burcsu, G. Ravichandran and K. Bhattacharya, Appl. Phys. Lett. **77** (11), 1698-1700 (2000).
29. S. Zhang, S.-M. Lee, D.-H. Kim, H.-Y. Lee and T. R. Shrout, Appl. Phys. Lett. **93** (12), 122908 (2008).
30. G. H. Jonker, J. Am. Ceram. Soc. **55** (1), 57-& (1972).
31. J. Deng, X. Ding, T. Lookman, T. Suzuki, K. Otsuka, J. Sun, A. Saxena and X. Ren, Phys. Rev. B **81** (22), 220101 (2010).
32. M. Kessel, R. A. De Souza and M. Martin, Phys. Chem. Chem. Phys. **17** (19), 12587-12597 (2015).
33. R. A. Maier and C. A. Randall, J. Am. Ceram. Soc. **99** (10), 3360-3366 (2016).